\documentclass[journal]{IEEEtran}
\ifCLASSINFOpdf
\else
\fi
\usepackage[pdftex]{graphicx}

\usepackage[numbers, sort&compress]{natbib}
\usepackage[caption=false,font=normalsize,labelfont=sf,textfont=sf]{subfig}
\usepackage[table]{xcolor}
\usepackage{amssymb}
\usepackage{float}
\usepackage{flushend}

\begin{document}
%
\title{AI-Empowered Channel Generation for IoV Semantic Communications in Dynamic Conditions} 
%
%
%

\author{Hao~Liu, Bo~Yang, Zhiwen Yu, Xuelin~Cao, George C. Alexandropoulos, 
Yan Zhang, and Chau Yuen
\thanks{


H. Liu and B. Yang are with the School of Computer Science, Northwestern Polytechnical University, Xi'an, Shaanxi, 710129, China (email: liuhao23@mail.nwpu.edu.cn, yang$\_$bo@nwpu.edu.cn). 

Z. Yu is with the School of Computer Science, Northwestern Polytechnical University, Xi'an, Shaanxi, 710129, China, and Harbin Engineering University, Harbin, Heilongjiang, 150001, China (email: zhiwenyu@nwpu.edu.cn).

X. Cao is with the School of Cyber Engineering, Xidian University, Xi'an, Shaanxi, 710071, China (email: caoxuelin@xidian.edu.cn). 

G. C. Alexandropoulos is with the Department of Informatics and Telecommunications, National and Kapodistrian University of Athens, 16122 Athens, Greece (email: alexandg@di.uoa.gr). 

Y. Zhang is with the Department of Informatics, University of Oslo, 0316 Oslo, Norway (email: anzhang@ieee.org).  

C. Yuen is with the School of Electrical and Electronics Engineering, Nanyang Technological University, Singapore (email: chau.yuen@ntu.edu.sg). 
 }
}


\maketitle

\begin{abstract}
The Internet of Vehicles (IoV) transforms the transportation ecosystem promising pervasive connectivity and data-driven approaches. Deep learning and generative Artificial Intelligence (AI) have the potential to significantly enhance the operation of applications within IoV by facilitating efficient decision-making and predictive capabilities, including intelligent navigation, vehicle safety monitoring, accident prevention, and intelligent traffic management. Nevertheless, efficiently transmitting and processing the massive volumes of data generated by the IoV in real-time remains a significant challenge, particularly in dynamic and unpredictable wireless channel conditions. To address these challenges, this paper proposes a semantic communication framework based on channel perception to improve the accuracy and efficiency of data transmission. The semantic communication model extracts and compresses the information to be transmitted. In addition, the wireless channel is estimated by using a generative diffusion model, which is employed to predict the dynamic channel states, thereby improving the quality of IoV service. In dynamic scenarios, however, the channel estimation performance may be degraded when substantially new scenarios take place, which will adversely affect user experience. To mitigate this limitation, we employ a large model to fine-tune the channel generation model to enhance its adaptability for varying scenarios. The performance and reliability of the proposed framework are evaluated on the two public datasets.
\end{abstract}

\begin{IEEEkeywords}
Semantic communications, Internet of Vehicles, Channel estimation, Generative diffusion model.
\end{IEEEkeywords}

\IEEEpeerreviewmaketitle

\section{Introduction}
\IEEEPARstart{I}{n} the realm of the Internet of Things (IoT), devices such as cameras, light detection and ranging (LiDAR), and other connected sensors facilitate multi-modal information transmission and fusion leveraging wireless communications.  In this context, the intelligent collaboration and data sharing among these devices significantly advance the development of the Internet of Vehicles (IoV). This paradigm enables the sharing of information, such as real-time location, vehicle speed, and road conditions, to optimize traffic management. Furthermore, it integrates multi-source data to provide users with intelligent navigation and early warning services, thereby enhancing both traffic safety and operational efficiency \cite{yb-magazine}.

Although IoV has achieved significant progress with the aid of Artificial Intelligence (AI) schemes, several challenges remain. For instance, the interconnected devices in the IoV generate vast amounts of data that require transmission and processing, which consequently increases processing delays in traditional communication networks. On another note, wireless bandwidth is often limited in the practical IoV scenarios, and it is essential to implement a low-latency and highly reliable data transmission method. To this end, Semantic Communications (SC) offer a promising solution by reducing the volume of data transmitted, thereby enhancing the transmission efficiency. 
This form of wireless communications pertains to the semantic level of communications, intending to accurately transmit the meaning underlying the message \cite{yang2022semantic}. An SC system usually consists of source encoders and decoders, channel encoders and decoders, and a shared knowledge base. To improve the robustness of this model and communication efficiency, source-channel encoders and decoders are usually jointly trained.

\begin{table*}[h]
    \centering
    \color{black}
    \caption{ \small Basic features of state-of-the-art approaches for semantic communication systems.}
    \label{related_work}
        \begin{tabular}{|l|l|p{2.2cm}|p{2.2cm}|p{2.4cm}|}
	\hline
		\multicolumn{1}{|l|}{Method}   & Efficient codec  & Low-rank devices adaptation    & Channel compensation   & Scene sensing-based model refinement    \\ \hline
		\multicolumn{1}{|l|}{CNN architecture \cite{bourtsoulatze2019deep,9746335}}   & \multicolumn{1}{c|}{\checkmark}  & \multicolumn{1}{c|}{\checkmark}  & \multicolumn{1}{c|}{\texttimes}  & \multicolumn{1}{c|}{\texttimes}   \\ \hline 
            \multicolumn{1}{|l|}{Transformer architecture \cite{10500305}}   & \multicolumn{1}{c|}{\checkmark}  & \multicolumn{1}{c|}{\texttimes}  & \multicolumn{1}{c|}{\texttimes}  & \multicolumn{1}{c|}{\texttimes}   \\ \hline 
		\multicolumn{1}{|l|}{Large model (e.g., GPT) \cite{xie2024towards,jiang2023large}}  & \multicolumn{1}{c|}{\checkmark}     & \multicolumn{1}{c|}{\texttimes}   & \multicolumn{1}{c|}{\checkmark}  & \multicolumn{1}{c|}{\texttimes} \\ \hline 
		\multicolumn{1}{|l|}{GAI (e.g., VAE, GAN, and diffusion) \cite{10447237,grassucci2023generative}}   & \multicolumn{1}{c|}{\checkmark}     & \multicolumn{1}{c|}{\checkmark}     & \multicolumn{1}{c|}{\texttimes}    & \multicolumn{1}{c|}{\texttimes}   \\ \hline
		\multicolumn{1}{|l|}{The proposed CTCD framework}   & \multicolumn{1}{c|}{\checkmark}     & \multicolumn{1}{c|}{\checkmark}   & \multicolumn{1}{c|}{\checkmark}  & \multicolumn{1}{c|}{\checkmark}  \\ \hline 
	\end{tabular}
\end{table*}
Advanced AI techniques have been recently increasingly applied to SC. After training, the SC model can reduce the amount of data transmitted, through data compression, to adapt to the constrained communication bandwidth. Since the data is interfered with by noise during the transmission process, which affects the user quality experience, the encoder-decoder architecture in SC systems plays a crucial role in maintaining the integrity and reliability of the transmitted content. Some popular SC models based on deep learning primarily utilize Convolutional Neural Networks (CNNs) \cite{bourtsoulatze2019deep,9746335} and a Transformer \cite{10500305} model. In addition, there exists recent interest in exploring the feasibility of applying large or generative models to SC systems. In fact, various foundational models, such as Generative Adversarial Networks (GANs) and Generative Diffusion Models (GDMs), have demonstrated strong capabilities in image and video generation. Large models have been successfully utilized for intent understanding and image generation \cite{xie2024towards}, and for training data generation to enhance the generalization of SC systems for dynamic scenarios. Due to their rich knowledge representation capabilities, large models can be integrated into the transceiver modules in SC systems. However, this integration imposes substantial demands on storage and computational resources \cite{xu2024unleashing}, making it difficult to deploy on resource-constrained edge or mobile devices.

In addition, in complex IoV environments, the highly dynamic and variable nature of wireless channels will make it difficult to predict and model the channel state accurately. Traditional methods struggle to capture real-time changes in the channel and often exhibit poor generalization performance. Consequently, an effective channel estimation method for IoV communications becomes necessary. To this end, Deep Learning (DL), compressed sensing, and online learning methods supporting the high accuracy channel estimation requirements of IoV systems have been recently designed. Existing DL-based approaches for channel estimation include CNNs~\cite{8640815}, Long Short-Term Memory (LSTM)~\cite{helmy2023lstm}, and Transformers~\cite{10529153}. However, these methods depend heavily on data quality and quantity, hence, their success in complex and time-varying channel conditions is highly questionable. 

Generative AI (GAI) techniques used in vision and language processing have showcased their profound advantages in data fitting, being capable to learn the data distribution via neural networks, or probabilistic generative models generating high-quality samples through sampling procedures, such as Variational AutoEncoders (VAE), GAN, and GDMs \cite{jiang2024largearXiv}.  
This indicates that GAI has the potential to facilitate the design, optimization, and management of IoV systems, where vehicle mobility is expected to lead to dynamic wireless channel characteristics. In this context, SC systems need to capture the dynamic changes of the channel in real-time, thereby supporting scenario-crossing channel modeling to improve the system's generalization. 

In this article, to deal with the dynamic channel conditions and the low-latency communication requirements in IoV systems, we present a novel SC framework that is based on a CNN-Transformer architecture \cite{zhang2023lite} offering channel estimation through a GDM; we abbreviate this framework as CTCD. In the proposed system, each image at the transmitter is first processed for semantic extraction and data compression, and the compressed data is transmitted over the wireless channel. To alleviate the impact of dynamic channel conditions, we employ a conditional diffusion network to construct a channel estimation model to generate Channel State Information (CSI). The received data undergo channel compensation followed by a channel-semantic decoder to reconstruct the original image. To adapt to dynamic channel environments, the channel model is fine-tuned using a large model. Furthermore, a route planning module is used to predict whether the vehicle moves into a new scene to determine whether the channel model needs fine-tuning. In the subsequent sections, we introduce the proposed CTCD framework in detail and conduct a series of experiments on public image datasets to verify its effectiveness. Table~\ref{related_work} summarizes the basic features of the state of the art in SC systems. in comparison with the proposed framework.

\section{Semantic Communications via GDM}
\subsection{Overview of the Proposed CTCD Architecture}
\begin{figure*}
    \centering
    \includegraphics[width=5.0in]{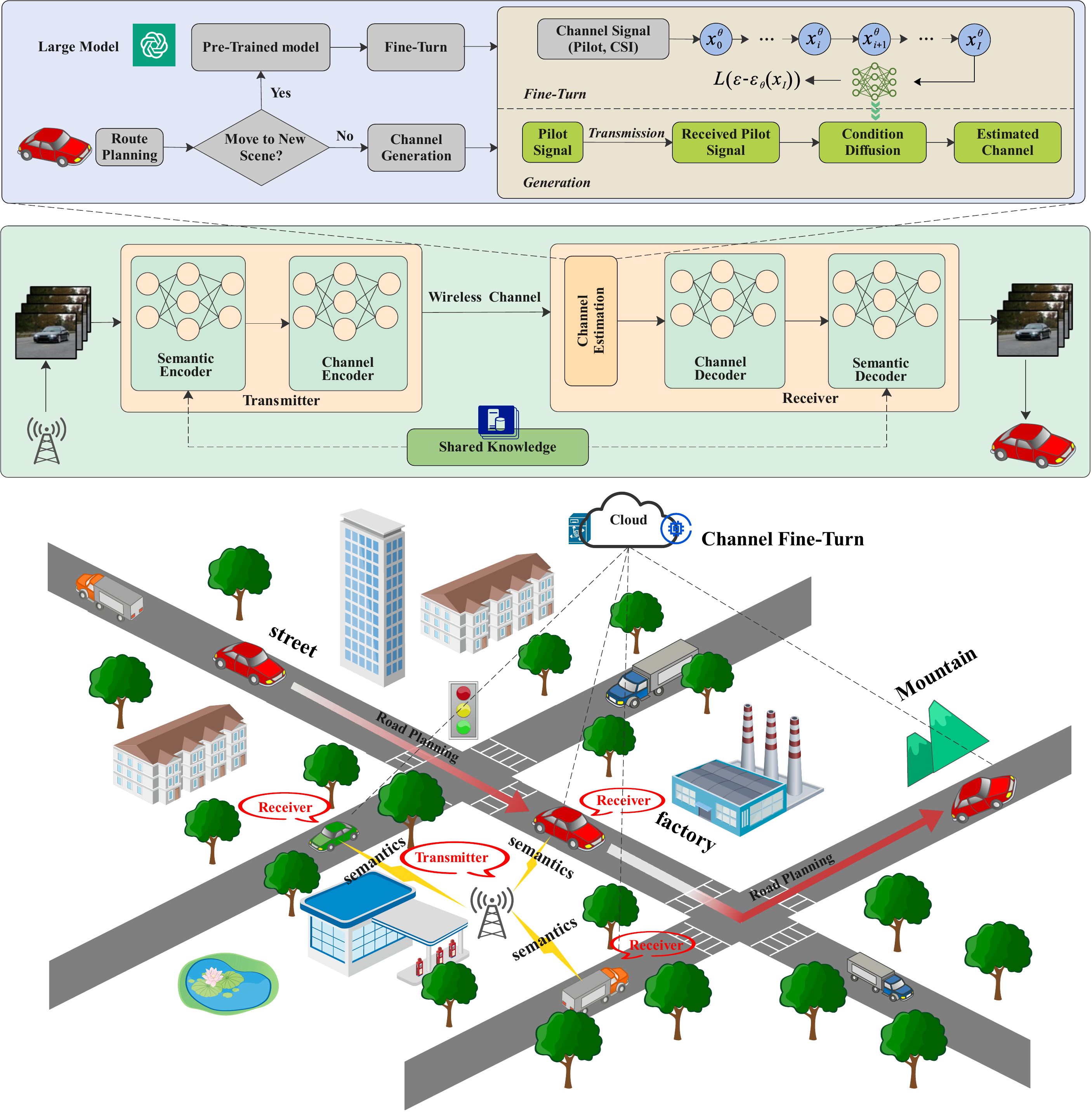}
    \caption{ \small The proposed AI-empowered framework for semantic communications.}
    \label{secenariodesign}
\end{figure*}

The proposed CTCD framework consists of three main modules: \textit{1)} the semantic information extraction and data recovery module; \textit{2)} a module for channel estimation via a conditional diffusion model; and \textit{3)} the channel refinement module that uses large models. The semantic extraction and recovery module adopts the CNN-Transformer architecture~\cite{zhang2023lite} as the semantic encoder-decoder to extract image semantic information and reconstruct images. Note that, compared with traditional communications, the channel has a greater impact on semantic communication. At the same bit error rate, data errors in semantic communication may destroy key semantic units. For example, in autonomous driving, errors of several bits may lead to incorrect target recognition, such as misidentifying pedestrians as vehicles. To consider realistic characteristics of wireless channels, we employ a conditional diffusion model to generate CSI to compensate for the interference of those channels on data. However, when a vehicle moves to a new scene, the performance of the channel estimation model may degrade. To deal with this issue, we first use vehicle route planning to obtain the vehicle's location. Then, we determine whether the vehicle appears in a new scene, and reserve sufficient time for the channel model refinement. By using a large model, the channel diffusion model can be fine-tuned in advance to improve the generalization in a different scene. The overall framework, components, and scenarios are illustrated in Fig.~\ref{secenariodesign}.

\subsection{CTCD Modules}
\subsubsection{Semantic Information Extraction and Data Recovery}
In general, a highly efficient source/channel encoder-decoder can extract more effectively semantic information and improve data reconstruction quality. As the depth and width of a DL model increase, the model's representation learning capabilities are enhanced. However, the storage and computing resources of the vehicles are usually limited due to constraints on the power supply (e.g., battery), so a lightweight model is desirable for practical deployment. Therefore, we use a lightweight visual model based on the CNN-Transformer architecture~\cite{zhang2023lite} to extract important semantic information from the image data, such as pedestrians and traffic signs. The neural network architecture has multiple stages, each containing three sublayers: a sampling layer, a CNN layer, and a transformer layer. The extracted semantic information includes both local and remote global information. The transformer layer is used to extract global information, and the dilated convolution used in the CNN layer expands the receptive field to extract local information without increasing the training parameters of the model, thereby meeting the resource-constrained demands in IoV applications.

During the data reconstruction process, the receiver first feeds the received data to the channel estimation module to perform channel compensation, and then uses the decoder to reconstruct the image. Note that the decoder and encoder have a symmetrical structure, and the generated image is provided to the user, or smart application, to understand the current road or traffic status, and thus make reasonable decisions.

\subsubsection{Channel Estimation via a Conditional Diffusion Model}
GAI schemes generate new samples leveraging the learning of the data distribution. Due to the complex distribution feature of wireless channels, traditional generative models, such as VAE and GAN, which are well suited for modeling low-dimensional data, struggle to produce high-quality samples. On the other hand, a diffusion model generates samples through a stepwise denoising process, hence, it can generate high-quality samples closer to the distribution of high-dimensional complex data. Therefore, in this article, we consider Denoising Diffusion Probabilistic Models (DDPM) for channel estimation. However, the preliminary diffusion generation process entirely depends on the noise and data distribution, and its randomness makes it difficult to generate accurate CSI. To this end, we use the conditionally guided diffusion model for channel estimation. In particular, the pilot signal is used as a condition to guide the model in adjusting the details of the sample during the generation process; this avoids generating unreasonable channel information. The generated CSI is then used to compensate for the interference caused by the real propagation.

\subsubsection{Channel Estimation Refinement with Large Models}
Data is disturbed by the wireless channel during transmission, which affects the data recovery process. The common practice is to estimate the wireless channel to compensate its effect at the receiver side. To this end, our SC system incorporates channel estimation. However, using a channel model trained offline to deal with this problem is challenging, since this approach lacks generalization. For example, when a vehicle moves to a new scene, adaptation to the propagation environment may be minimal, resulting in performance degradation. To alleviate this problem, our framework deploys a large model which is hosted in the cloud to refine the channel diffusion model, and thus improve its generalization. The large model is trained with channel data from many scenarios that contain different wireless propagation features, thereby achieving stronger generalization and adaptation ability to handle complex scenarios. 
We have particularly used the GPT-2 as our large model architecture. 

For the refinement step, we have mainly focused on minimizing the computational time. Overall, the required time consists of the data transmission time, the large model inference time, and the channel model refinement time. Recall that, to achieve channel compensation upon entering a new scene, the vehicles needs to initiate the fine-tune process in advance. To support this, we have utilized vehicle route planning to determine whether refinement is necessarym and when the fine-tuning process should start.  
Specifically, by inputting the coordinates of the current position and the target position into the map, the vehicle's driving route can be obtained. Along this route, geographic coordinates are sampled at each distance (e.g., every kilometer), and then it is determined whether the coordinate is located in a new scene. If the future coordinate is in a new scene, the system calculates the optimal starting point for refinement based on the total time required for the process, thereby ensuring the diffusion model can be updated before the vehicle arrives at that new scene.

\section{Case Study: Image Transmission}
\begin{figure}[!t] 
    \centering 
        \includegraphics[width=3.35in]{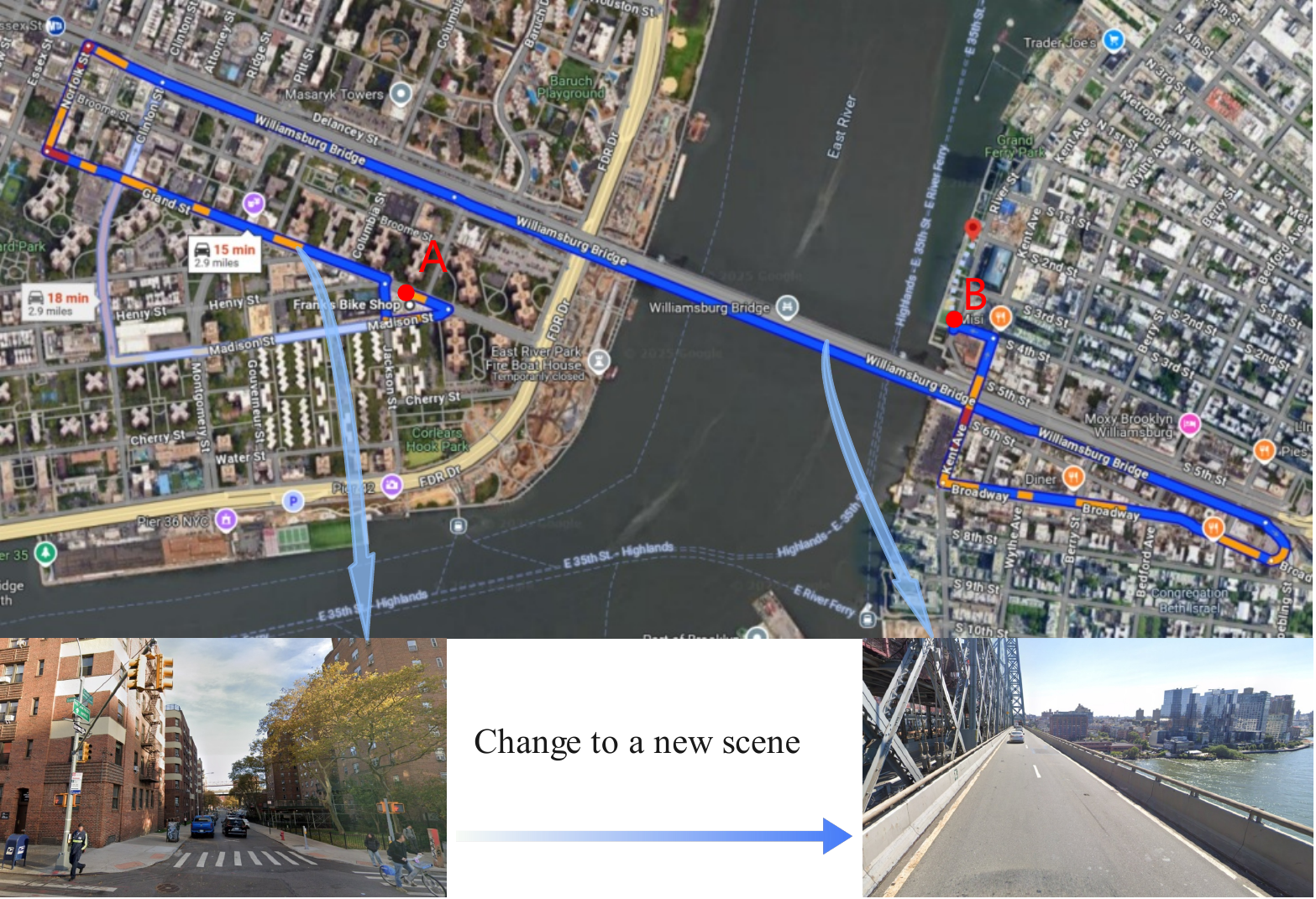}
    \caption{ \small The considered case study: a vehicle moves in New York city and its driving route involves different scenarios with various channel propagation conditions. }
    \label{map_route_and_scenario_change}
\end{figure}

In this section, we investigate the performance of the proposed CTCD framework for an image transmission task, e.g., pedestrian detection and accident warning, which is typical in IoV scenarios. The real-time traffic data transmission requires low latency but high reliability communications to ensure traffic safety and improve user experience. We have considered a vehicle in New York City as a case study, where the communication occurs from the base station the vehicle or directly (Vehicle-to-vehicle (V2V) communications). As shown in Fig.~\ref{map_route_and_scenario_change}, the vehicle moves from point A to point B. According to the route planning, the vehicle passes through a city road scenario and then enters a cross-river bridge scenario, which is relatively different. During this whole process, the vehicle obtains services through our CTCD framework.

\begin{figure*}[!t] 
    \centering
   
        \subfloat[]{\includegraphics[width=1.7in]{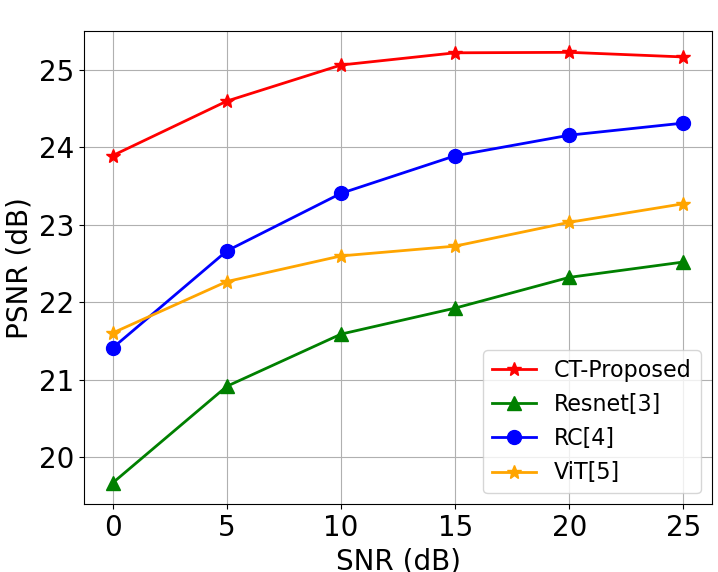}
        \label{stanford_cars_psnr}} 
        \subfloat[]{\includegraphics[width=1.74in]{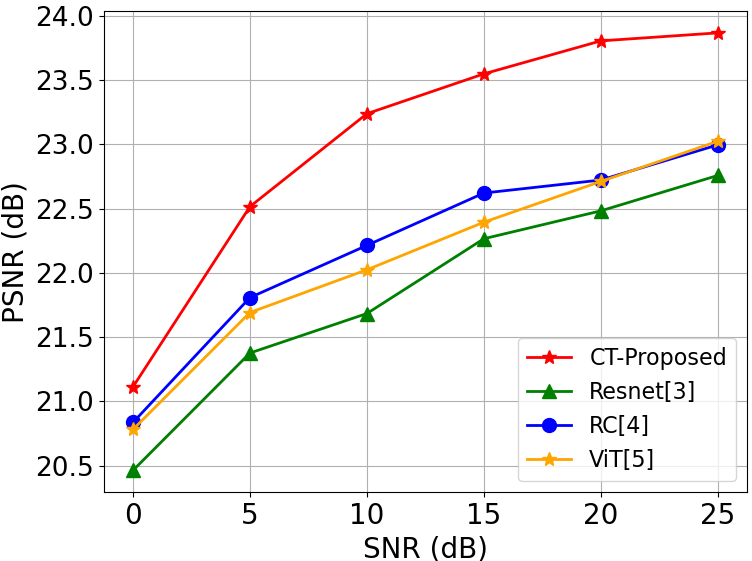}
        \label{trafficnet_dataset_v1_psnr_compare}} 
        \subfloat[]{\includegraphics[width=1.84in]{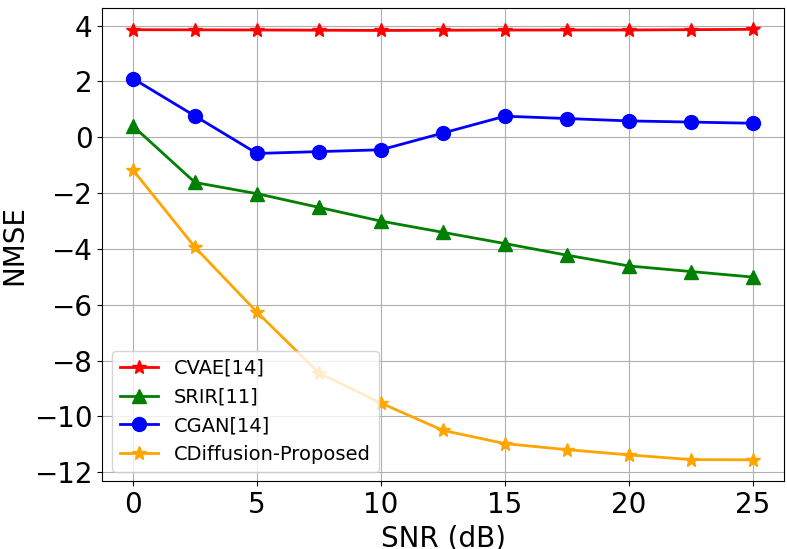}%
        \label{Channel_estimate_NMSE}}  
        \subfloat[]{\includegraphics[width=1.89in]{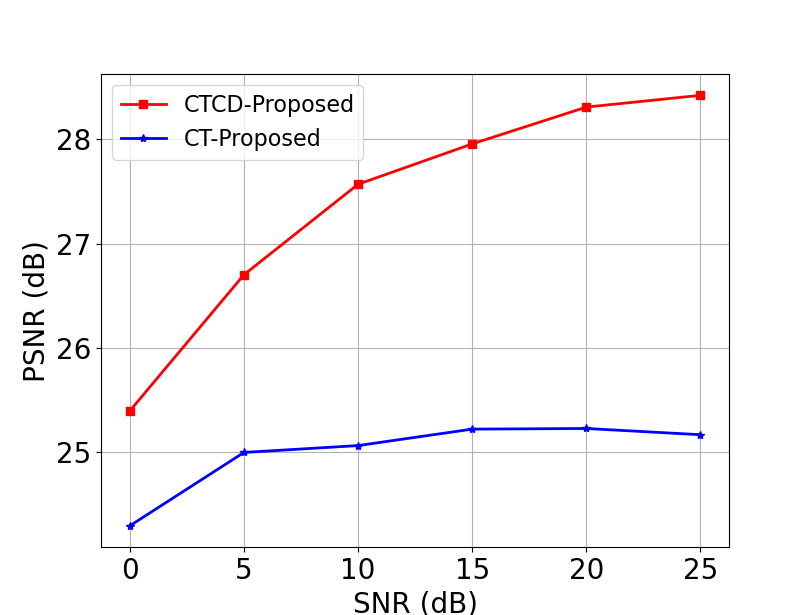}%
        \label{stanford_cars_with_channel_model}}
    \caption{ \small Communication and channel estimation performance. PSNR versus the SNR for the (a) Stanford\_cars and (b) Traffic-Net datasets is depicted for different SC frameworks. The NMSE performance versus the SNR is illustrated in (c) for various channel estimation models. (d) PSNR versus the SNR considering SC with and without channel estimation over the Stanford\_Cars dataset. (Legends: CT-Proposed denotes this articles' SC framework that uses the CNN-Transformer~\cite{zhang2023lite} as its semantic codec; Resnet is~\cite{bourtsoulatze2019deep}'s model; RC denotes Rate-Control SC with CNN~\cite{9746335}; ViT is the Vision Transformer~\cite{10500305}; CDiffusion-Proposed represents Channel estimation with a Diffusion model in our framework; and CTCD-Proposed is our overall proposed SC framework. In addition, for the channel estimation process, CVAE and CGAN stand for the Conditional versions of VAE and GAN in~\cite{jiang2024largearXiv}; and SRIR is \cite{8640815}'s Super Resolution and Image Restoration model.)}
    \label{psnr_three_datasets} 
\end{figure*}

To verify the effectiveness and evaluate the performance of the proposed framework, a series of experiments were conducted on the public IoV image datasets Stanford\_Cars and Traffic-Net. The datasets were used for the SC task of image reconstruction, hence, we used an image quality evaluation metric, specifically, the Peak Signal-to-Noise Ratio (PSNR). Moreover, we considered the Mean Squared Error (MSE) function to jointly train the source-channel encoder and decoder, and the performance of the our framework was tested at multiple Signal-to-Noise Ratio (SNR) values. Our SC framework has been compared with the following DL-based baselines for semantic information encoding and decoding: Residual Network (Resnet)~\cite{bourtsoulatze2019deep}, Vision Transformer (ViT) \cite{10500305}, and RC \cite{9746335}, where RC is a CNN model-based policy network with balanced rate and feature quality. Besides, we have utilized a conditional diffusion model to generate CSI to compensate for the impact of the channel on the data. VAE, GAN, as well as Super-Resolution CNN (SRCNN) and Denoising CNN (DnCNN) were the other state-of-the-art models on channel estimation used for comparison.  

Figure~\ref{psnr_three_datasets} compares the proposed CTCD framework with the aforementioned baseline schemes. As observed from Fig. \ref{psnr_three_datasets}(a) and~(b), the PSNR performance increases with increasing SNR. Compared with the state-of-the-art models, it is shown that the proposed CNN-Transformer (CT) achieves better results when used as an encoder-decoder, showcasing that CT can combine the advantages of CNN and Transformer architectures. By extracting effective semantic information through CT, image data can be better reconstructed when subject to the wireless channel. However, the VIT architecture has more parameters than the CNN architecture, and the higher computational complexity of the multi-head attention in this transformer leads to slower model inference. 

Unlike traditional communications, data extraction and compression in SC lead to high density information, which significantly amplifies the system's sensitivity to channel conditions. We have designed a Channel estimation module through the conditional Diffusion model (CDiffusion) to compensate for the impact of the channel, the performance of which have been tested over the DeepMIMO dataset, which includes a variety of wireless communication scenarios (e.g., indoor, outdoor, garage, and multiple urban scenarios). For comparison purposes, three baselines were implemented, as shown in~Fig. \ref{psnr_three_datasets}(c). We first studied the models' training and evaluation when the scene does not change, considering the Normalized MSE (NMSE) metric. It can be observed that, as the SNR increases, the MMSE of CDiffusion (that generates CSI) decreases, indicating the effectiveness of this model in learning the distribution of the wireless channel, outperforming all benchmarks. Finally, Fig.~\ref{psnr_three_datasets}(d) showcases our CTCD framework's PSNR performance for the Stanford\_Cars dataset in comparison with CT, verifying the efficacy of our joint design of the CT and CDiffusion modules.
\begin{figure}[!t] 
    \centering
        \subfloat[]{\includegraphics[width=1.67in]{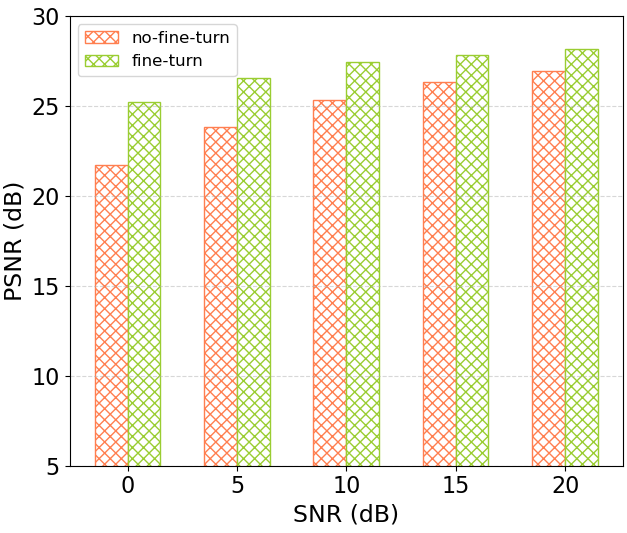}%
        \label{channel_fineturn_psnr_compare}} 
        \subfloat[]{\includegraphics[width=1.78in]{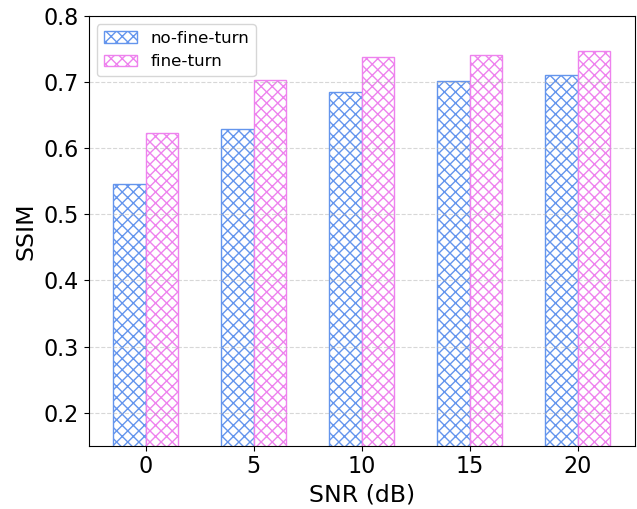}%
        \label{channel_fineturn_ssim_compare}}
    \caption{ \small PSNR and SSIM versus the SNR considering the proposed large model for channel estimation refinement upon the Stanford\_Cars dataset for the case where the vehicle moves between scenes with different channel conditions. Respective results for the case without channel estimation fine-tuning have been also included.}
    \label{Channel_estimate_fine_turn}
\end{figure}

\begin{figure*}[!t]
    \centering
    \includegraphics[width=7.1in]{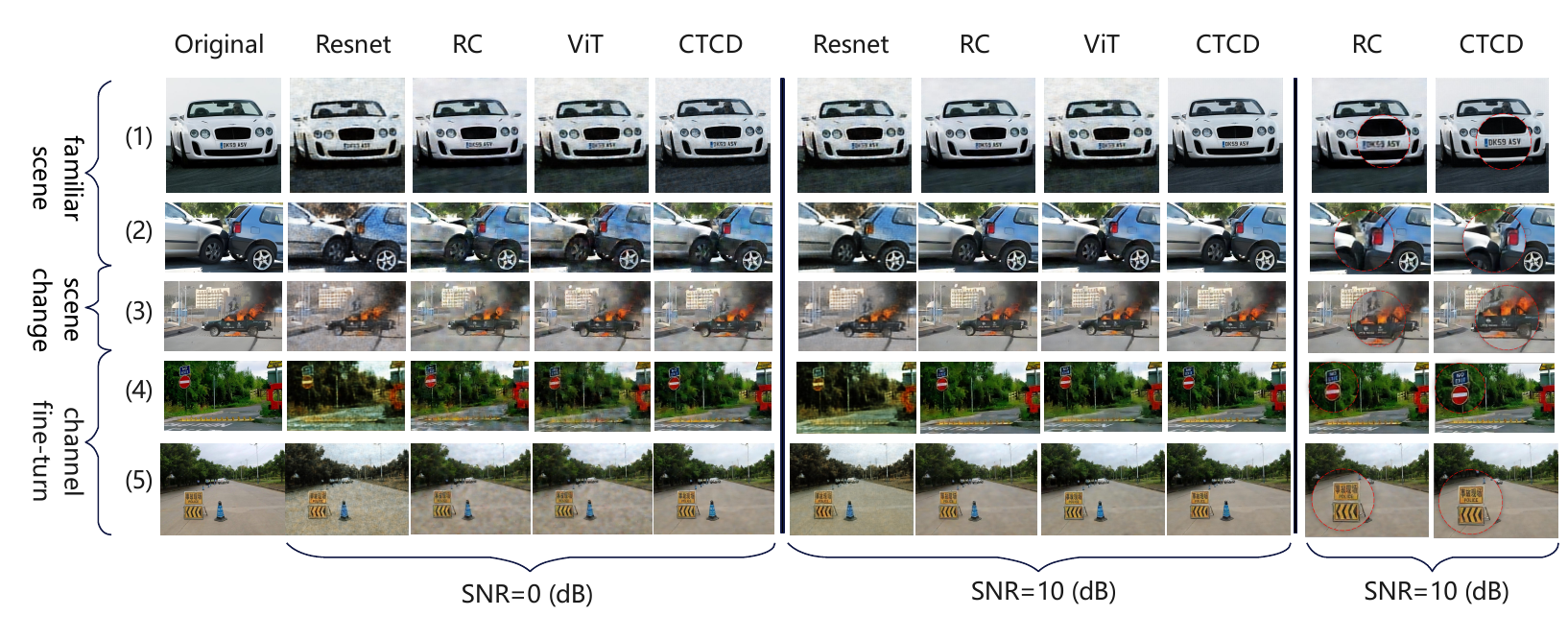}
    \caption{ \small The received images of five different IoV scenes for two SNR values using SC with the proposed CTCD model and the benchmark models Resnet, RC, and ViT. The two right-hand columns include the results for the two best schemes focusing on the critical parts of all five images (red circles). The different rows indicate whether there has been a scene change between the training and deployment phases, and whether the proposed large-model-based channel estimation refinement has been used with the proposed CTCD model.}
    \label{ImageVisualization}
\end{figure*}

As previously discussed, in mobile vehicle networks, when the vehicle moves to a new scene, the channel model may have difficulty to adapt to it, thus affecting the quality of the SC system. We have used GPT2 as the large model architecture and trained it using the communication data available in the DeepMIMO dataset. Then, the trained large model was employed to refine the GDM for channel estimation. We have evaluated the effectiveness of this approach by comparing the performance of SC systems when the scene changes with and without fine-tuning in Fig.~\ref{Channel_estimate_fine_turn}. It can be seen that the proposed refinement boosts both the PSINR and the SSIM metrics, especially when the operating SNR is low.

The communication quality of different IoV scenes (car detection, traffic accident, car fire, traffic sign, and road works) using the proposed CTCD model and the benchmark models Resnet, RC, and ViT for two SNR values is illustrated in Fig.~\ref{ImageVisualization}. The image sample in the first row was chosen from the Stanford\_cars dataset, having the highest relatively resolution, while the rest four images were selected from the Traffic-Net dataset and have lower resolution. The visualization results (1) and (2) were taken considering familiar scenes between model training and deployment. In the results (3), we have considered a scene change for all models, and the proposed CTCD model was used without the channel estimation refinement model. For the results (4) and (5), CTCD has been applied together with the proposed large-model-based channel estimation fine-tuning mechanism. It can be observed that, among the compared baselines, RC exhibits the best image restoration for both the SNR values $0$ and $10$ dB. In the two right-hand columns, the results with RC and CTCD at $10$~dB of SNR are compared focusing on the critical parts of all five images (in red circles). As depicted, the perspicuity with the proposed scheme is clearer, showcasing the superiority of CTCD over all benchmarks. It can be also seen from the results in rows (4) and (5) that, by using the proposed channel refinement model, image restoration becomes clearer, thus, validating its efficacy for improved data reconstruction.

\section{Conclusion and Future Directions}
In this article, we introduced a novel GDM-based SC scheme for efficient image transmission in IoV applications. The proposed framework leverages the CNN-Transformer architecture to extract image semantic information, and deploys a large model on a cloud server to refine local channel diffusion models, aiming at enhanced generalization when scenes change. The main challenges resulting from the proposed SC framework, which constitute interesting directions for future research, are the following.

\subsection{Energy Management}

Although an SC system requires less energy for transmissions, it needs to implement neural networks and semantic parsing algorithms, which significantly increase the computational energy consumption. In some cases, the energy used for processing can surpass the energy savings from the reduced amount of transmissions. This is more pronounced when fine-tuning dynamic channel models. 
For autonomous driving applications, self-driving vehicles must maintain an end-to-end latency, usually less than $100$ milliseconds. This requirement compels electronic control units to infer results in nearly real-time. Therefore, power saving becomes essential and can be achieved through edge AI, lightweight models, and sustainable energy management schemes. These strategies aim to provide continuous services for AI-enabled SC vehicular networks.

\subsection{Resource Allocation}

When designing conventional communication resource allocation schemes, it is essential to consider both quality of service and quality of experience. However, SC systems do not focus on the individual bits transmitted. This implies that conventional resource allocation algorithms need to be re-designed. For resource allocation dealing with image transmission, more bandwidth needs to be allocated to the data containing richer semantic information. One potential solution could be to integrate a bandwidth allocation awareness module into the SC system, which would be jointly trained with the rest of the models. This approach would allow for dynamic bandwidth allocation enabling the system to adjust the data compression rate according to the available bandwidth. Additionally, more energy should be allocated for transmitting data with higher semantic content to enhance energy efficiency.

Our SC framework incorporates a large model, hosted on a cloud server(s), to refine the local channel model. Vehicle movement may cause fluctuations in communication bandwidth, which can interrupt the fine-tuning process, thus, negatively impacting channel model parameter updates. In such cases, the fine-tuning of the channel model will require significant computational resources, which combined with the limited computing power available in vehicles may lead to halting other high-priority tasks, such as collision warnings. 

The latter issues imply that semantic information needs to be wisely utilized to jointly optimize communication and computing resources in future SC systems.

\subsection{Semantic Consistency}

In the IoV paradigm, it is essential to establish a uniform understanding of the semantics of the dynamic wireless environment and to facilitate conflict-free decision-making among distributed vehicular nodes. All vehicles in the network should provide semantic descriptions of the same event or object, such as ``road traffic accident ahead'' or ``intention to change lanes by the vehicle on the left,'' that are logically compatible and align spatially and temporally. This alignment will help preventing collaborative failures that can arise from semantic ambiguities or contradictions. Additionally, vehicles need to exchange semantic information in real-time, but the communication channels between them can vary. This variation may degrade the performance of the channel model when multiple vehicles engage in SC. Therefore, it is crucial to improve the generalization of the channel model, since failure to do so may lead to inconsistent semantic understanding between the sender and the receiver. A key area of study is related to the design of effective knowledge management schemes leveraging federated learning to enhance global models for vehicles. This would involve constraining local semantic generation to be consistent with global standards, thus ensuring coherent communication across the IoV.



\ifCLASSOPTIONcaptionsoff
  \newpage
\fi



%

\footnotesize
\bibliographystyle{IEEEtran}
\bibliography{IEEEexample}

\begin{thebibliography}{10}
\providecommand{\url}[1]{#1}
\csname url@samestyle\endcsname
\providecommand{\newblock}{\relax}
\providecommand{\bibinfo}[2]{#2}
\providecommand{\BIBentrySTDinterwordspacing}{\spaceskip=0pt\relax}
\providecommand{\BIBentryALTinterwordstretchfactor}{4}
\providecommand{\BIBentryALTinterwordspacing}{\spaceskip=\fontdimen2\font plus
\BIBentryALTinterwordstretchfactor\fontdimen3\font minus \fontdimen4\font\relax}
\providecommand{\BIBforeignlanguage}[2]{{%
\expandafter\ifx\csname l@#1\endcsname\relax
\typeout{** WARNING: IEEEtran.bst: No hyphenation pattern has been}%
\typeout{** loaded for the language `#1'. Using the pattern for}%
\typeout{** the default language instead.}%
\else
\language=\csname l@#1\endcsname
\fi
#2}}
\providecommand{\BIBdecl}{\relax}
\BIBdecl

\bibitem{yb-magazine}
B.~Yang, X.~Cao, K.~Xiong, C.~Yuen, Y.~L. Guan, S.~Leng, L.~Qian, and Z.~Han, ``Edge intelligence for autonomous driving in {6G} wireless system: Design challenges and solutions,'' \emph{IEEE Wireless Commun.}, vol.~28, no.~2, pp. 40--47, 2021.

\bibitem{yang2022semantic}
W.~Yang, H.~Du, Z.~Q. Liew, W.~Y.~B. Lim, Z.~Xiong, D.~Niyato, X.~Chi, X.~Shen, and C.~Miao, ``Semantic communications for future internet: Fundamentals, applications, and challenges,'' \emph{IEEE Commun. Surveys \& Tuts.}, vol.~25, no.~1, pp. 213--250, 2022.

\bibitem{bourtsoulatze2019deep}
E.~Bourtsoulatze, D.~B. Kurka, and D.~G{\"u}nd{\"u}z, ``Deep joint source-channel coding for wireless image transmission,'' \emph{IEEE Trans. Cogn. Commun. Netw.}, vol.~5, no.~3, pp. 567--579, 2019.

\bibitem{9746335}
M.~Yang and H.-S. Kim, ``Deep joint source-channel coding for wireless image transmission with adaptive rate control,'' in \emph{Proc. IEEE ICASSP}, Singapore, 2022, pp. 5193--5197.

\bibitem{10500305}
H.~Wu, Y.~Shao, E.~Ozfatura, K.~Mikolajczyk, and D.~Gündüz, ``Transformer-aided wireless image transmission with channel feedback,'' \emph{IEEE Tran. Wireless Commun.}, vol.~23, no.~9, pp. 11\,904--11\,919, 2024.

\bibitem{xie2024towards}
H.~Xie, Z.~Qin, X.~Tao, and Z.~Han, ``Towards intelligent communications: Large model empowered semantic communications,'' \emph{arXiv preprint arXiv:2402.13073}, 2024.

\bibitem{jiang2023large}
F.~Jiang, L.~Dong, Y.~Peng, K.~Wang, K.~Yang, C.~Pan, and X.~You, ``Large {AI} model empowered multimodal semantic communications,'' \emph{arXiv preprint arXiv:2309.01249}, 2023.

\bibitem{10447237}
H.~Du, G.~Liu, D.~Niyato, J.~Zhang, J.~Kang, Z.~Xiong, B.~Ai, and D.~I. Kim, ``Generative {AI}-aided joint training-free secure semantic communications via multi-modal prompts,'' in \emph{Proc. IEEE ICASSP}, Seoul, South Korea, 2024, pp. 12\,896--12\,900.

\bibitem{grassucci2023generative}
E.~Grassucci, S.~Barbarossa, and D.~Comminiello, ``Generative semantic communication: Diffusion models beyond bit recovery,'' \emph{arXiv preprint arXiv:2306.04321}, 2023.

\bibitem{xu2024unleashing}
M.~Xu, H.~Du, D.~Niyato, J.~Kang, Z.~Xiong, S.~Mao, Z.~Han, A.~Jamalipour, D.~I. Kim, X.~Shen \emph{et~al.}, ``Unleashing the power of edge-cloud generative {AI} in mobile networks: A survey of {AIGC} services,'' \emph{IEEE Commun. Surveys \& Tuts.}, vol.~26, no.~2, pp. 1127--1170, 2024.

\bibitem{8640815}
M.~Soltani, V.~Pourahmadi, A.~Mirzaei, and H.~Sheikhzadeh, ``Deep learning-based channel estimation,'' \emph{IEEE Commun. Lett.}, vol.~23, no.~4, pp. 652--655, 2019.

\bibitem{helmy2023lstm}
I.~Helmy, P.~Tarafder, and W.~Choi, ``{LSTM-GRU} model-based channel prediction for one-bit massive {MIMO} system,'' \emph{IEEE Trans. Veh. Technol.}, vol.~72, no.~8, pp. 11\,053--11\,057, 2023.

\bibitem{10529153}
S.~Zhang, S.~Zhang, Y.~Mao, L.~K. Yeung, B.~Clerckx, and T.~Q.~S. Quek, ``Transformer-based channel prediction for rate-splitting multiple access-enabled vehicle-to-everything communication,'' \emph{IEEE Trans. Wireless Commun.}, vol.~23, no.~10, pp. 12\,717--12\,730, 2024.

\bibitem{jiang2024largearXiv}
F.~Jiang, Y.~Peng, L.~Dong, K.~Wang, K.~Yang, C.~Pan, and X.~You, ``Large generative model assisted {3D} semantic communication,'' \emph{arXiv preprint arXiv:2403.05783}, 2024.

\bibitem{zhang2023lite}
N.~Zhang, F.~Nex, G.~Vosselman, and N.~Kerle, ``Lite-mono: A lightweight cnn and transformer architecture for self-supervised monocular depth estimation,'' in \emph{Proc. IEEE/CVF Conf. Computer Vision Pattern Recogn.}, Vancouver, Canada, 2023, pp. 18\,537--18\,546.

\end{thebibliography}

\begin{IEEEbiographynophoto}{Hao Liu (IEEE Student Member)} is currently pursuing a Doctoral Degree in the School of Computer Science, Northwestern Polytechnical University, Xi'an, Shaanxi, 710129, China.
\end{IEEEbiographynophoto}
\vskip -2\baselineskip plus -1fill

\begin{IEEEbiographynophoto}{Bo Yang (IEEE Member)} is a Professor in the School of Computer Science, Northwestern Polytechnical University, Xi'an, Shaanxi, 710129, China. 
\end{IEEEbiographynophoto}
\vskip -2\baselineskip plus -1fill

\begin{IEEEbiographynophoto}{Zhiwen Yu (IEEE Senior Member)} is a Professor in the School of Computer Science, Northwestern Polytechnical University, Xi'an, Shaanxi, 710129, China, and also with Harbin Engineering University, Harbin, Heilongjiang, 150001, China. 
\end{IEEEbiographynophoto}
\vskip -2\baselineskip plus -1fill

\begin{IEEEbiographynophoto}{Xuelin Cao (IEEE Member)} is an Associate Professor in the School of Cyber Engineering, Xidian University, Xi'an, Shaanxi, 710071, China. 
\end{IEEEbiographynophoto}
\vskip -2\baselineskip plus -1fill

\begin{IEEEbiographynophoto}{George C. Alexandropoulos (IEEE Senior Member)} 
is an Associate Professor in the Department of Informatics and Telecommunications, National and Kapodistrian University
of Athens, Greece  and an Adjunct Professor with the Department of Electrical and Computer Engineering, University of Illinois Chicago, USA. His research interests span the general areas of algorithmic design and performance analysis for wireless networks with emphasis on multi-antenna transceiver hardware architectures, full duplex MIMO, active and passive multi-functional RISs, ISAC, and millimeter-wave/THz
communications, as well as distributed machine learning algorithms. 
\end{IEEEbiographynophoto}
\vskip -2\baselineskip plus -1fill

\begin{IEEEbiographynophoto}{Yan Zhang (IEEE Fellow)} is currently a Full
Professor with the Department of Informatics,
University of Oslo, Norway. He is Fellow of IEEE, Fellow of IET, elected member of Academia Europaea (MAE), elected member of the Royal Norwegian Society of Sciences and Letters (DKNVS), and elected member of
Norwegian Academy of Technological Sciences (NTVA).
\end{IEEEbiographynophoto}
\vskip -2\baselineskip plus -1fill

\begin{IEEEbiographynophoto}{Chau Yuen (IEEE Fellow)} is an Associate Professor in the School of Electrical and Electronics Engineering, Nanyang Technological University, Singapore. 
\end{IEEEbiographynophoto}

\end{document}